\documentclass[aps,pra,reprint,superscriptaddress,amsmath,amssymb]{revtex4-2}

\usepackage{graphicx,mathtools,epstopdf,amssymb,amsmath,commath,verbatim,csquotes,ulem}
\usepackage{pifont}
\usepackage{tikz}
\newcommand*\circled[1]{\raisebox{.5pt}{\textcircled{\raisebox{-.9pt} {#1}}}}
\newcommand{\PT}{\mathcal{PT}}
        
\DeclareMathOperator{\sech}{sech}

\DeclareMathOperator{\sgn}{sgn}
\usepackage{color}

\usepackage[colorlinks=true,bookmarks=true,allcolors=blue,citecolor=red,urlcolor=blue]{hyperref}

\begin{document}

\title{Radiation families emitted by a discrete soliton in parity-time-symmetric waveguide arrays}
	
\author{Anuj P. Lara}
\affiliation{Department of Physics, Indian Institute of Technology Kharagpur, Kharagpur 721302, India}
\author{Ambaresh Sahoo}
\affiliation{Department of Physical and Chemical Sciences, University of L’Aquila, Via Vetoio, L’Aquila 67100, Italy}
\author{Samudra Roy}
\email{samudra.roy@phy.iitkgp.ac.in}
\affiliation{Department of Physics, Indian Institute of Technology Kharagpur, Kharagpur 721302, India}

\begin{abstract}

We investigate the dynamics of a spatial discrete soliton and the radiation families emitted by it inside a parity-time ($\mathcal{PT}$)-symmetric waveguide array with alternate gain-loss channels. A strong spatial soliton that evolves inside the waveguide array due to the balance between discrete diffraction and Kerr nonlinearity excites linear waves in the form of diffractive radiation when launched with an angle. $\mathcal{PT}$-symmetric nature of the waveguide leads to additional radiations in Fourier space that were never explored before. In our work, we mainly focus on the origin of these radiations and try to understand how to control them. Under strong  $\mathcal{PT}$ symmetry, a discrete soliton launched normally to the waveguide array produces strong side-lobes which can lead to a population of field at $\pm \pi/2$ in momentum space. In addition, a strong soliton with initial phase gradient radiates unique $\mathcal{PT}$ symmetry assisted linear wave. We establish a phase matching condition to locate such radiation in momentum space. The periodic arrangement of the gain-loss channel also leads to radiations due to reflection and back-scattering, which is prominent for a weak soliton. A linear Hamiltonian analysis for such a waveguide array is provided to identify the $\mathcal{PT}$-phase transition regime and to optimize the parameter for stable discrete soliton dynamics. We thoroughly investigate the origin of all the radiations that emerged in the $\mathcal{PT}$-symmetric waveguide array and put forward the background theory which is in good agreement with the full numerical results.

\end{abstract}

\maketitle
	
\section{Introduction}
A linear waveguide array (WA) is a periodic photonic structure where a propagating optical wave experiencing spatially periodic refractive index distribution behaves like an electron traveling through a semiconductor crystal. The 1D homogeneous WAs that are evanescently coupled to each other allow the light beam to propagate in a transverse direction, resulting in a phenomenon called \textit{discrete diffraction}. The optical Kerr nonlinearity counterbalances the discrete diffraction and exhibits localized structure in the form of \textit{discrete soliton} (DS). The dynamics of DS is governed by the discrete nonlinear Schrödinger equation (DNLSE). These spatially localized structures manifest properties that are intriguing and forbidden in the case of their continuous counterpart. Few such examples are, Anderson localization \cite{lahini_anderson_2008,martin_anderson_2011}, photonic Bloch oscillations, localized Wannier-Stark states \cite{morandotti_experimental_1999, pertsch_optical_1999}, Bloch-Zener oscillation and photonic Zener tunneling \cite{breid_blochzener_2006, dreisow_bloch-zener_2009}. 

Diffractive resonant radiation (DifRR), which is the spatial (or wavenumber) analog of the dispersive resonant radiation in the time (frequency) domain, has primarily been investigated in uniform WAs supporting DS as a theoretical perspective \cite{tran_diffractive_2013}.
This concept is further extended in supercontinuum generation in both frequency and the wavenumber domains\cite{tran_supercontinuum_2014}.
Unlike temporal dispersive radiations, which are controlled by zero-dispersion wavelengths and span over any frequency range \cite{skryabin_colloquium_2010}, the DifRR strictly confines within the first Brillouin boundary ($\pm\pi$) in wavenumber space. Any electric field that encounters this Brillouin boundary results in $2\pi$ phase-shift and  appear from the opposite wavenumber boundary, a phenomenon termed as {\it anomalous recoil} \cite{tran_diffractive_2013}.  In uniform WAs, the generation of DifRR requires an initial nonzero wavenumber of DS, while its manipulation and control can be achieved through the soliton power and coupling. However, with the introduction of chirp in WAs \cite{lara_dynamic_2020}, there is an additional degree of freedom (chirp) other than the wavenumber that tailors the generation of DifRR. In this scheme, it is even possible to generate dual DifRR with the introduction of a symmetrically chirped WA.

Moving forward with the WAs, in recent years the non-Hermitian Hamiltonian with $\PT$ symmetry or broadly speaking non-Hermitian quantum physics has emerged as a fascinating topic in both theoretical and experimental physics after Bender and Boettcher put forward the mathematical idea in 1998 \cite{bender_real_1998}. 
A non-Hermitian Hamiltonian $\hat{H}$ is considered to be $\PT$-symmetric if $[\hat{H},\PT]=0$, where $\mathcal{P}$ and $\mathcal{T}$ respectively denote the parity (space reflection) and time-reversal ($i=\sqrt{-1}$ flips sign) operators. One notable aspect of such Hamiltonians is the breaking of $\PT$ symmetry, in which the eigenspectra transition from completely real (unbroken $\PT$ regime) to complex values (broken $\PT$ regime) across a $\PT$ phase transition point called \textit{exceptional point} (EP), where atleast two eigenvalues and corresponding eigenvectors coalesce and become degenerate \cite{miri_exceptional_2019,el-ganainy_non-hermitian_2018}.
The concept of $\PT$ symmetry is majorly appreciated in the field of optics which offers a suitable platform  to practically demonstrate the unique non-Hermitian features such as unidirectional light propagation or optical non-reciprocity \cite{peng_paritytime-symmetric_2014}, loss-induced transparency \cite{guo_observation_2009}, optical solitons \cite{wimmer_observation_2015}, enhanced light-matter interactions \cite{wiersig_enhancing_2014,hodaei_enhanced_2017}, etc.

$\PT$-symmetric WAs with balanced loss/gain offer new possibilities in shaping light beams, which have been used in various contexts \cite{christodoulides_discrete_1988,Kartashov_2019,Zhu_2020}. In this work, we explore the generation of new kinds of radiations  in momentum space ($k$-space) of WAs with alternate gain and loss channels supporting DSs. With the introduction of $\PT$-symmetric WAs, we observe two new families of radiation, assisted solely by the $\PT$ symmetry. The locations of these radiations can be controlled further by changing the gain/loss coefficient that opens up new possibilities in designing novel photonic devices and controlling the flow of light at the nanoscale level.

We organize the paper as follows. In section II we describe the set-up and establish the governing equation for a discrete soliton excited in a  $\mathcal{PT}$-symmetric WA.  We also provide a background theory of DifRR and develop a linear Hamiltonian analysis for $\mathcal{N}$- channel  $\mathcal{PT}$-symmetric WA. In section III, we carefully analyze different radiations emitted by the DS in $k$-space due to the $\mathcal{PT}$-symmetric nature of the WA. Finally in section IV we summarize our results and conclude.

\section{System Set Up and Theoretical Framework}
\begin{figure}[h]
\includegraphics[width=\linewidth]{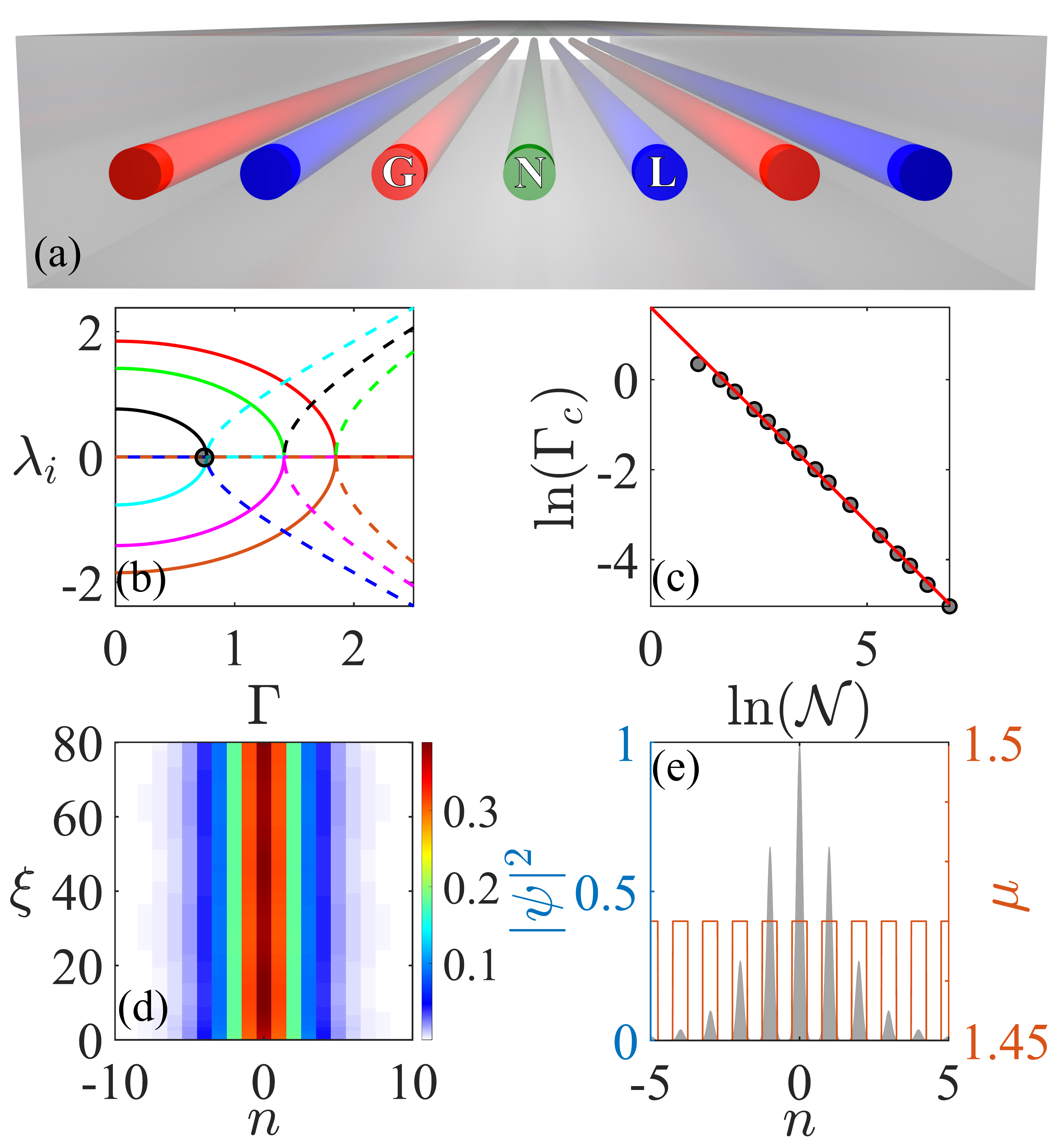}
	\caption{ (a) Schematic diagram for the gain and loss profile of a $\mathcal{PT}$-Symmetric waveguide array, with a neutral waveguide as a central defect.
		(b) The variation of real (solid) and imaginary (dotted) components of the eigen values of a 7 channel $\mathcal{PT}$-Symmetric WA as a function of $\Gamma$. Here, circle represents the exceptional point ($\Gamma_c$).
		(c) Variation of the critical point (exceptional point) $\Gamma_{\rm c}$ as a function of the total number of waveguides $\mathcal{N}$ (in natural log scale).
	(d) A discrete soliton in a waveguide array, with (e) the mode distribution along with the refractive index profile.
	}
\label{fig:system}
\end{figure}

To begin with the $\PT$-symmetric waveguide setup that excites DS, we consider an array of optical waveguides that are weakly coupled to their adjacent neighbors with linear gain/loss and are arranged in an alternate layout with the central waveguide ($n = 0$) being a neutral waveguide, as schematically shown in  \figref{fig:system} (a). We use a minimalistic and simple design here by keeping the spacing between two adjacent waveguides constant over the entire array. Assuming that the coupling between gain-loss, gain-neutral, and neutral-loss channels are all identical, the mode evolution in the $n$th waveguide $E_n$ can be described by the following normalized equation. 
\begin{equation}
i \frac{d \psi_n}{d \xi} + c [\psi_{n+1} + \psi_{n-1}] n + |\psi_n|^2 \psi_n =  \sgn(n)(-1)^n i \Gamma \psi_n,
\label{eq:primary}
\end{equation}
where $\psi_n (\xi,\tau)=E_n/\sqrt{P_0}$, with $P_0$ being the peak power of the input field. Range of the index $n$ is considered within $-N \leqslant n \leqslant N$, thereby defining the total number of waveguides to be $\mathcal{N}=(2N + 1)$. The propagation distance ($z$) and coupling coefficient ($C$) between adjacent waveguides are rescaled as $\xi \rightarrow z/L_{\rm NL}$ and $c \rightarrow CL_{\rm NL}$, where nonlinear length is defined by $L_{\rm NL}=(P_0 \gamma)^{-1}$ and $\gamma$ is the nonlinear parameter in units of W$^{-1}$m$^{-1}$. The gain/loss parameter $g$ having the unit of m$^{-1}$ is rescaled as $\Gamma=gL_{\rm NL}$.
Since this paper aims to investigate the formation and control of DifRRs in $\PT$-symmetric WAs, first we study the dynamics of DSs in the simplest $\PT$-symmetric WAs as described in \figref{fig:system} (a). We excite DSs in the unbroken $\PT$-regime  since the broken $\PT$-regime causes instability \cite{wimmer_observation_2015}.  To identify the unbroken $\PT$-regime,  we implement a Hamiltonian analysis and determine the parameter space (relation between the coupling and gain/loss coefficients) for stable DS excitation. 

\subsection{Hamiltonian Analysis}
For the $\mathcal{N}$-channel $\PT$-symmetric WA with central defect (neutral channel) (see \figref{fig:system}(a)), one can construct a $\PT$-symmetric Hamiltonian $\hat{H}$ with unitary coupling ($c=1$) and balanced gain/loss $\Gamma$, which satisfies $[\hat{H},\PT]=0$, as
\begin{equation}
\hat{H} \equiv 
        \begin{bmatrix}
		-i \Gamma &  1 &  0 & . & . & . &. &. &0\\
		1 &  i \Gamma & 1 &0 &. &. &. &. &0\\
		0 & 1 & -i \Gamma & 1 & 0 &. &. &. &0\\
		. & . &. & . &. &. &. & .&  .\\
		0 & . &. & .& . & . & . & . & 0\\
		. & . & . & . & . & . & . & . &.\\
		0 & . & . & . & . & . & 0 &1 & (-1)^\mathcal{N} i \Gamma
        \end{bmatrix}_{\mathcal{N} \times \mathcal{N}} \label{eq:N_Hamil}.
\end{equation}
The $\mathcal{N}$ eigenvalues of the system can be obtained by diagonalizing the Hamiltonian matrix. 
Considering a three-channel gain-neutral-loss waveguides system for simplicity, the [$3 \times 3$] Hamiltonian can be written as $\hat{H}_3 \equiv [-i \Gamma,\, 1,\, 0; \,1,\, 0,\, 1;\,0 ,\,1,\, i \Gamma]$, which results in the eigenvalues 0 and $\pm \sqrt{2 - \Gamma^2}.$
Here, $\Gamma = \Gamma_c = \sqrt{2}$ acts as the critical value at which the eigenvalues transition from real to imaginary with the real components collapsing to zero at this point \cite{hodaei_enhanced_2017,sahoo_two-way_2022}.
This critical point is known as an EP, where the $\mathcal{PT}$ symmetry breaking takes place (a third-order EP in this case).
For higher number of waveguide channels the eigenvalues take more complex forms and need to be evaluated numerically.
Such a set of eigenvalues are obtained numerically for a system of 7-channels and plotted in \figref{fig:system} (b). Note, the lowest value of $\Gamma_c$ depends on the number of waveguide channels. To establish the empirical relation between waveguide channel ($\mathcal{N}$) and $\Gamma_c$ we numerically calculate the set of  eigenvalues from the [$\mathcal{N} \times \mathcal{N}$] Hamiltonian as given in Eq. \eqref{eq:N_Hamil}. In  \figref{fig:system} (c) , we plot  $\Gamma_{\rm c}$ as a function of $\mathcal{N}$ in the logarithmic scale showing a linear dependency as $\ln(\Gamma_{\rm c}) \approx  A \ln(\mathcal{N}) + B$, where $A=-0.954$ and $B= 1.606$. This empirical relation allow us to determine the threshold value for unbroken  $\mathcal{PT}$-regime to excite stable DS.

\subsection{Discrete soliton and diffractive radiation }	
	
In absence of any $\mathcal{PT}$-symmetric potential ($\Gamma=0$) the propagation equation becomes, 	
	\begin{align}
	i \frac{d \psi_n}{d \xi} + c\left[\psi_{n+1} + \psi_{n-1}\right] + \abs{\psi_n}^2 \psi_n = 0,
	\label{eq:nDNLSE}
\end{align}	
which is the standard form of DNLSE. In absence of nonlinearity ($\gamma=0$) Eq.\eqref{eq:nDNLSE}	reduces to an analytically integrable equation whose solution $\psi_n(\xi)=\psi_n(0)i^nJ_n(2c\xi)$ exhibits discrete diffraction. 
Now exploiting the discrete plane-wave solution $\psi_n(\xi) = \psi_0\left[i(nk_xd + \beta \xi)\right]$ of Eq.\eqref{eq:nDNLSE}, one can obtain the standard dispersion relation between the longitudinal
wave vector  $\beta$ and $k_x$ as	$\beta(\kappa) = 2c\cos(\kappa) + \abs{\psi_0}^2,$
where $d$ is the separation between two adjacent waveguides, $k_x$ is the transverse wave vector and $\kappa \equiv k_x d$ is the Bloch-momentum defined as phase difference between two adjacent waveguides \cite{christodoulides_discrete_1988}.
The Taylor expansion of $\beta(\kappa)$ about the incident wavenumber ($\kappa_0$) results in the diffraction relation:
\begin{align}
	\beta(\kappa) = \beta(\kappa_0) + \sum\limits_{m\geqslant1}\frac{D_m}{m!}\Delta\kappa^m,
	\label{eq:diffac_eqn}
\end{align}
where $D_m \equiv \left(d^m \beta/d \kappa^m\right)|_{\kappa_0}$ and $\Delta \kappa = \kappa -\kappa_0$. The parameter $D_1$ represents the transverse velocity.
Performing a Fourier transformation to change the domain as $\kappa \rightarrow n$ by replacing $\Delta \kappa \equiv -i \partial_n$, where $n$ is defined as a continuous variable of an amplitude function $\Psi(n,\xi)=\psi_{n,\xi}\exp(-i\kappa_0n)$, we obtain an approximate standard \textit{nonlinear Schr\"{o}dinger equation}(NLSE)\cite{tran_diffractive_2013}
\begin{equation}
	\left[i \partial_{\xi}  + \sum\limits_{m\geqslant2} \frac{D_m}{m!}\left(-i \partial_n\right)^m + \abs{\psi(n,\xi)}^2\right]\psi(n,\xi) = 0.	
	\label{eq:sNLSE}
\end{equation}
 One can eliminate  the first and second term of the Taylor expansion by making a transformation $\psi(n,\xi) \rightarrow \psi(n,\xi)\exp\left[i \beta(\kappa_0) n\right]$ and considering the co-moving frame $n\rightarrow (n+D_1 \xi$).
 For $D_{m\geqslant3} = 0$ the exact solution of  Eq.\eqref{eq:sNLSE} is given as,
$\psi_{sol} = \psi_0 \sech\left(\frac{n \psi_0}{\sqrt{\abs{D_2}}}\right) \exp\left(i k_{sol} \xi\right),$
	 here $k_{sol} \equiv \psi_0^2/2$ is the longitudinal wavenumber of the soliton.
Note that a bright soliton exists only when the condition $\abs{\kappa_0} < \pi/2$ or $2c\cos(\kappa_0)>0$ is satisfied.
\figref{fig:system}(d) describes formation of such  soliton that propagates in nonlinear uniform WA for an input beam $\psi_{sol}=\psi_0 \sech(n\psi_0/\sqrt{|D_2|})$. In \figref{fig:system}(e) the spatial distribution of the  DS is  illustrated in the background of periodic refractive index grid offered by the typical WA. Note, here the evolving soliton encompass several waveguides which justifies the continuous variable approximation of $n$. The plane-wave solution  $\exp\left[i\left(k_{lin}\xi + \Delta \kappa n\right)\right]$ of the linearized Eq.\eqref{eq:sNLSE} leads to the dispersion relation	$	k_{lin}(\Delta \kappa) = \beta(\kappa)- \beta(\kappa_0) - D_1 \Delta \kappa$. A soliton  emits  radiation in $\kappa$-space by transferring energy to the linear wave when the condition $k_{sol} = k_{lin}(\Delta \kappa)$ is satisfied for a specific $\Delta k$. The phase matching (PM) condition for generating DifRR \cite{tran_diffractive_2013}, can be written as,
	\begin{align}
	\left[\cos(\kappa) - \cos(\kappa_0) + \sin(\kappa_0)\Delta \kappa\right] = \widetilde{\psi_0}^2
	\label{eq:PM}
\end{align}
where $\widetilde{\psi_0} = \psi_0/2\sqrt{c}$.	

\begin{figure}[]
	\begin{center}
		\includegraphics[width=\linewidth]{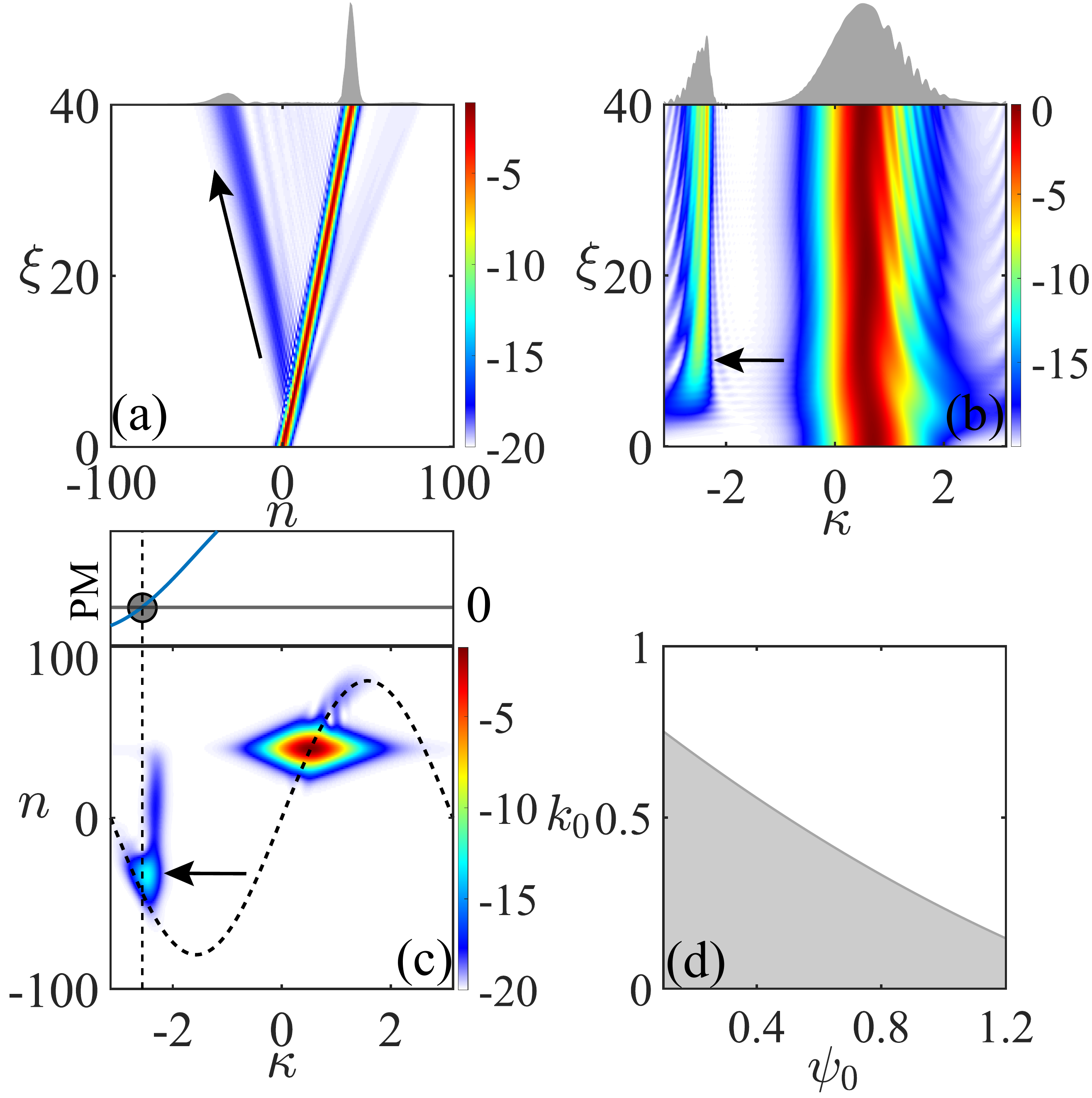}
	\end{center}
	\caption{(a) Formation of the weak linear wave corresponding to DifRR in $n$-space when a tilted DS is launched. (b) A prominent signature of DifRR in $\kappa$-space around the value $\kappa=-2.8$  and (c) the corresponding  XFROG spectrogram at output (marked by the arrow). The dotted line represents the spatial delay line defined as $\delta_s=D_1\xi$. In the inset we plot the PM expression [Eq.\eqref{eq:PM}] whose solution predicts the Bloch-momentum ($\kappa_{RR}$) of DifRR. (d) Phase plot showing the regime of ($k_0$,$\psi_0$) values to generate the DifRR. The gray region is forbidden for DifRR to generate. }\label{fig:2}
\end{figure}
The solution of Eq.\eqref{eq:PM} gives the wavenumber of the generated DifRR ($\kappa_{RR} = \kappa_0 + \Delta\kappa$) as a function of the initial soliton wavenumber $\kappa_0$ which is related to the incident angle of the beam.
In \figref{fig:2}(a), formation of the DifRR in $n$-space is demonstrated when the DS is launched with an inclination.
The signature of  DifRR (around $-2.8$) is prominent in $\kappa$-space as illustrated in \figref{fig:2}(b). In \figref{fig:2} (c) we plot the spectrogram which is mathematically defined as, $\mathcal{S}(n,\kappa,\xi)=|\int_{-\infty}^{\infty} \psi(n',\xi)\psi_{w}(n-n')e^{i\kappa n'}dn'|^2$, where $\psi_{w}$ is a reference window function (normally taken as input). This plotting scheme allows us to represent the output in $n$-$\kappa$ space where  we clearly observe the  formation of DifRR and  DS. We calculate the spatial delay line $\delta_s=D_1\xi$ at output which is analogous to the temporal delay usually calculated for the supercontinuum generation process in photonic crystal fibers. This delay line indicates the relative locations of different components (soliton and radiations) in $n$-$\kappa$ space. Note, in the reduced scheme, the formation of DifRR is restricted within the first Brillouin zone ($\kappa_{RR} < \abs\pi$) and undergo a shift (formally knows as \textit{anomalous recoil}) in it's wavenumber by $\pm 2\pi$ when they form outside these limits. In the top inset of \figref{fig:2} (c) we plot the PM expression [Eq.\eqref{eq:PM}] whose solution predicts the Bloch-momentum ($\kappa_{RR}$) of DifRR. The formation of DifRR is found to be sensitive to the relative values of $k_0$ and $\psi_0$. In \figref{fig:2} (d) we provide a phase-plot in $k_0$-$\psi_0$ space showing a forbidden region (gray region) for DifRR to generate. That means DifRR  are suppressed for the set of ($k_0$,$\psi_0$) values falling on the gray region.

	\section{Diffractive radiation in $\mathcal{PT}$-symmetric waveguide array}
	
\begin{figure}[h]
	\begin{center}
		\includegraphics[width=\linewidth]{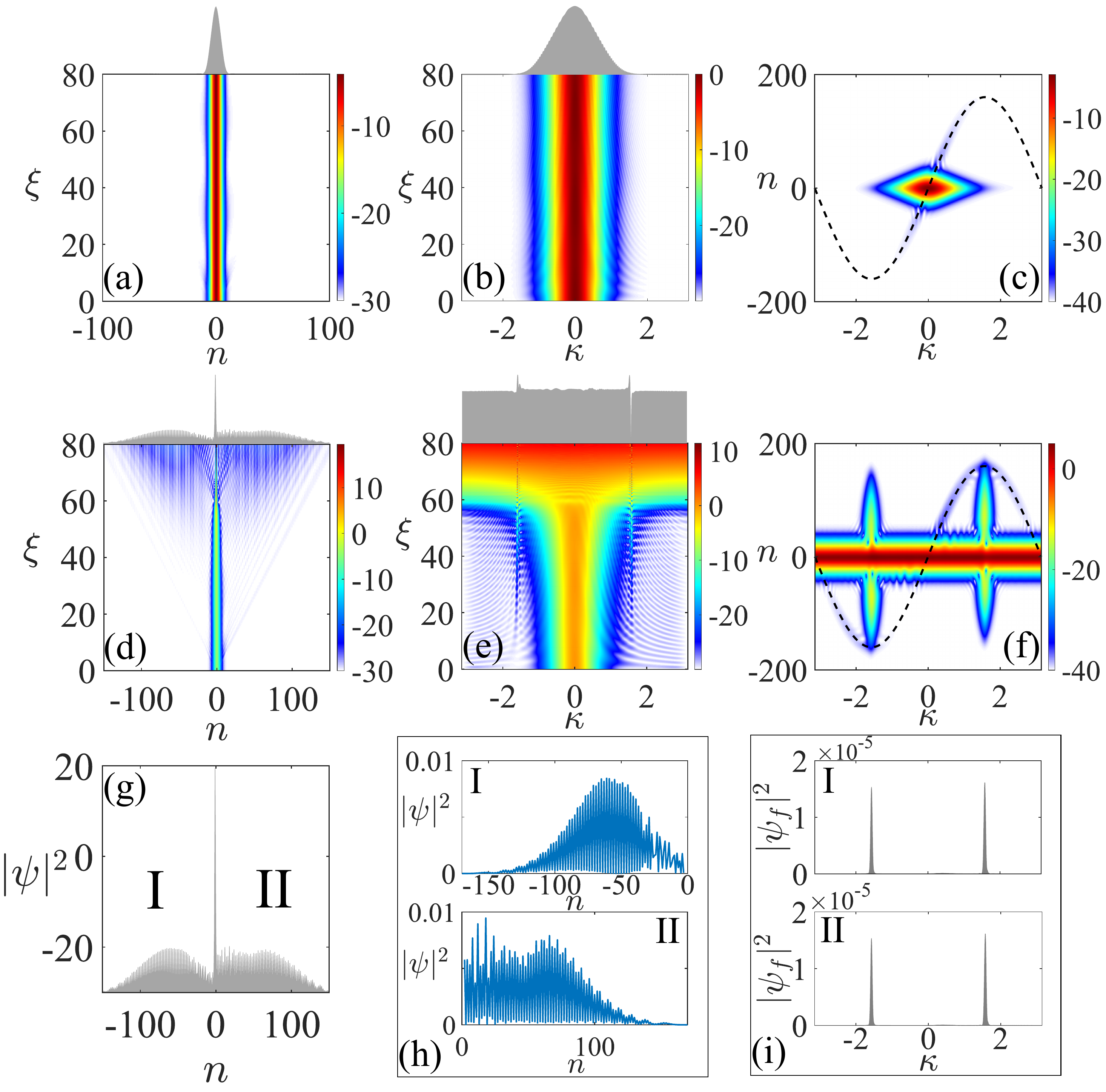}
	\end{center}
	\caption{ Spatial soliton formation in (a) $n$-space and (b) $\kappa$-space with (c) spectrogram at the output $\xi = 80$. Here soliton amplitude $\psi_0 = 0.8$ and $\Gamma$ of $0.01$ ($<\Gamma_c$). Dynamics of DS in (d) $n$-space and (e) $\kappa$-space with (f) spectrogram at the output $\xi = 80$ for $\Gamma = 0.08$  ($>\Gamma_c$) in $\PT$-symmetry breaking regime. (g) Tight confinement of soliton for $\Gamma$ beyond EP exhibiting strong side-lobes represented as I and II. (h) Individual representation of the side-lobes (I = top, II = bottom) and (i) their Fourier transform showing radiations at $\kappa = \pm \pi/2$.}	
	\label{fig:3}
\end{figure}

In this section we investigate the dynamics of a DS in semi-infinite  $\mathcal{PT}$-symmetric waveguide array with a neutral central waveguide. The schematic diagram of the WA is depicted in  \figref{fig:system} (a), where asymmetric gain/loss channels are arranged in either sides of the central waveguide ($n=0$). 
 For a large number of waveguides ($\mathcal{N}$), EP approximately obeys the relation, $\ln(\Gamma_{\rm c}) \approx A \ln(\mathcal{N}) + B$.
For a linear $\mathcal{PT}$-symmetric WA with $\mathcal{N}=501$ the  EP reduces to as small as $\Gamma_c \approx 0.013$. In \figref{fig:3} we compare the dynamics of DS for unbroken and broken symmetry regime by taking two  different $\Gamma$ values.  In  \figref{fig:3}(a)-(c) we demonstrate the evolution of a DS launched at the neutral waveguide ($n=0$) for $\Gamma=0.01$ (which is in unbroken symmetry regime $\Gamma<\Gamma_c$). In this limit we observe  propagation of a stable soliton which  spreads over a few waveguide channel. The field is also localized in the Fourier domain [$\kappa$-domain, see \figref{fig:3}(b)]. The spectrogram in \figref{fig:3}(c) clearly indicates the formation of a localized field at output.   Dramatic change is observed when we increase the $\mathcal{PT}$ parameter to  $\Gamma=0.08$ ($> \Gamma_c$) which corresponds to the  broken symmetry regime. In this limit the soliton which was initially extended to a few waveguides, squeezes to a single waveguide (preferably the wavegude channel with gain) as shown in  \figref{fig:3}(d).  This tight confinement of the soliton  in $n$-domain results in an outburst in the Fourier space ($\kappa$-domain) showing a continuous band [see  \figref{fig:3}(e) and (f)]. 
Side-lobes are formed in $n$ channels as the energy is flowing through the assistance of alternative gain channels of the WA. This field distribution of side-lobes  results in two peaks in the Fourier domain ($\kappa$-space) at precise location  $\pm \pi/2$ evident in \figref{fig:3}(f). 
The side wings I and II in \figref{fig:3}(g) independently contribute  to the peaks. A Fourier transform of the two side lobes I and II (plot (h)) results in two distinct peaks at $\pm \pi/2$ [plot (i)].
These are the signature peaks for a $\mathcal{PT}$-symmetric waveguide array when a soliton is excited beyond EP.

\subsection{Dynamics of strong DS (for $k_0 \neq 0$) in $\mathcal{PT}$-symmetric WA}

\begin{figure}[h]
	\begin{center}
		\includegraphics[width=\linewidth]{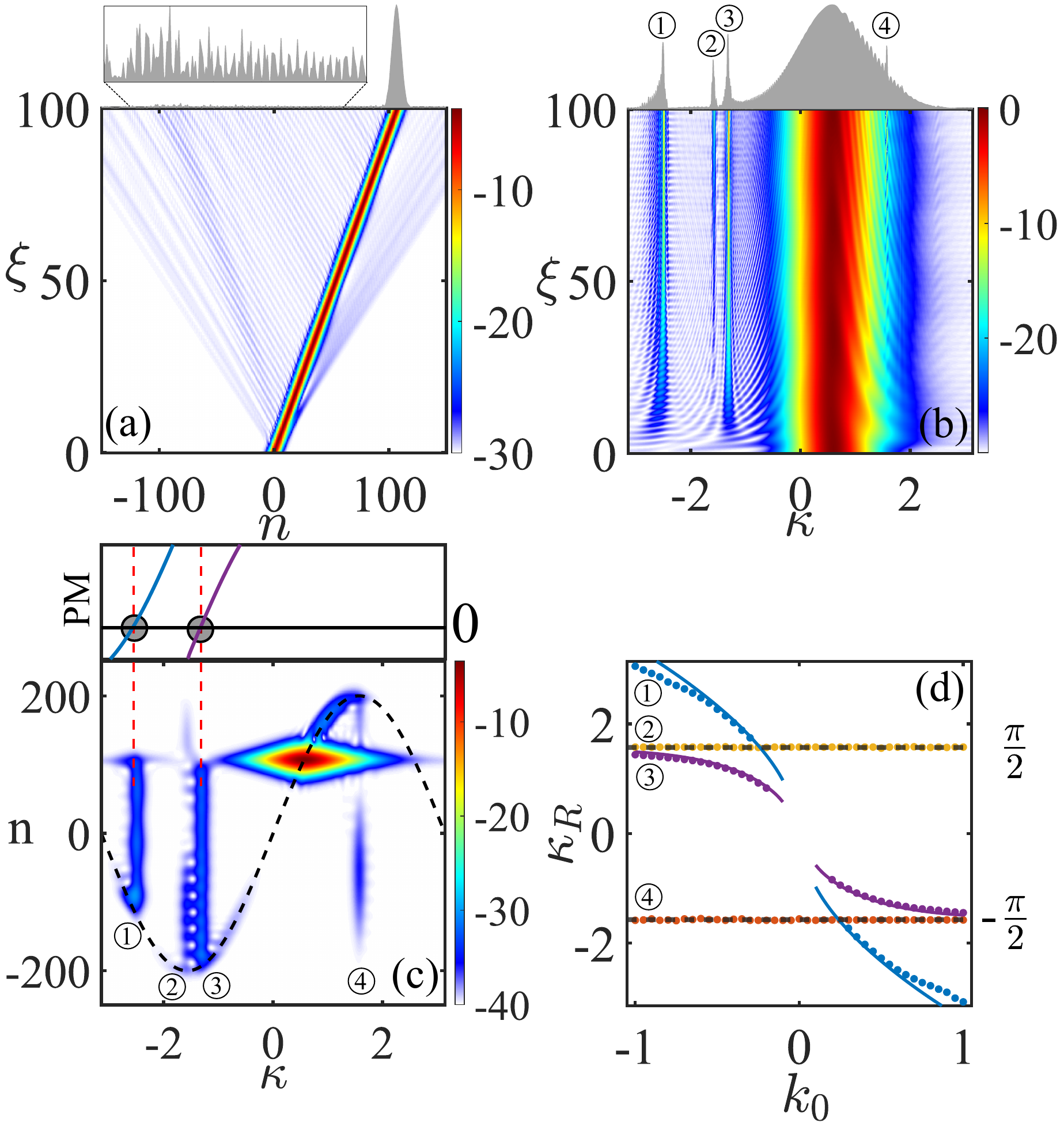}
	\end{center}
	\caption{(a) Soliton propagation with initial amplitude $\psi_0 = 0.6$ and wavenumber $k_0 = 0.6$ for $\Gamma = 0.04$. In the inset we magnify the localization of the $\mathcal{PT}$-symmetry assisted pedestal field. 
		  (b) Emissions of different radiations in $\kappa$-space marked as \circled{1}- \circled{4}.
		(c) Spectrogram at the output. In the inset we plot  phase matching (PM) relation for both the standard DifRR \circled{1} and a $\mathcal{PT}$ symmetry assisted radiation \circled{3}. 
		(d) Variation of DifRR \circled{1} and $\mathcal{PT}$ symmetry assisted radiation \circled{3} as a function of $k_0$, where numerical results (solid dots) are compared with proposed PM equations (solid lines). 
		 \circled{2} and \circled{4} indicate the characteristic radiation at $\pm \pi/2$ (black dashed lines) independent of $k_0$.}
	\label{fig:4}
\end{figure}

In this sub-section we concentrate on the dynamics of a strong DS inside a  $\mathcal{PT}$-symmetric WA and try to investigate the background mechanism of radiations emitted by this strong soliton.
It is already  demonstrated that DifRRs are emerged over a threshold value of the pair ($\psi_0$, $k_0$). Generally for strong solitons (with high $\psi_0$) the formation of DifRR is evident.  
Here a  strong soliton ($\psi_0=0.6$) is launched with a phase gradient (non-zero $k_0$ value) as shown in \figref{fig:4} (a). This propagation dynamics leads to  four distinct radiation peaks in $\kappa$-space marked by \circled{1}-\circled{4}  in  \figref{fig:4} (b). The radiation marked as \circled{1} is distinctly identified as DifRR. However, the origin of other three radiations is solely due to the $\mathcal{PT}$-symmetric characteristics of WA because they are absent when the $\mathcal{PT}$ symmetry is switched off. 
\circled{2} and \circled{4} represent the characteristic radiations at $\pm \pi/2$ generated due to the $\mathcal{PT}$ symmetry assisted side-lobe propagation. It is worthy to note, for a weak $\Gamma$ value these radiations are suppressed.
Therefore, in the simulation we consider a value of $\Gamma$ above the threshold EP value of the system such that soliton collapse does not happen up to a reasonable distance and at the same time the $\pm \pi/2$ radiations are prominent. 
In the inset of \figref{fig:4} (a) we magnify the pedestal part of the moving soliton which shows how the side-lobes are confined preferentially in alternative gain  channels resulting a periodic localized distribution in $n$-space. A simple spatial  Fourier transform of this periodic distribution of the pedestal field leads to the distinct radiation  at  $\pm \pi/2$ in $\kappa$-space.
Interestingly, in the Fourier space we observe another radiation marked as \circled{3} whose origin needs to be addressed. To understand the complete picture we plot the spectrogram  in \figref{fig:4} (c), where all the radiations appeared as stains in $n$-$\kappa$ space over spatial delay line $\delta_s$ (dotted curve). It is observed that, the radiations \circled{3} and \circled{2} are closely spaced in $\kappa$-space and spread over in  $n$- domain. Unlike \circled{2} and \circled{4} the location of the radiation \circled{3} are found to be sensitive to the initial launching angle ($k_0$ value) of the DS.
In  \figref{fig:4} (d) we track down all the radiations as a function of $k_0$ ranging from $-1$ to $+1$. Note that for $|k_0|<0.25$ the radiations \circled{1} and \circled{3} are suppressed.
However, as expected, the  characteristic radiations  \circled{2}, \circled{4} located at $\pm \pi/2$  remain unaffected by the variation of $k_0$.  While the phase matching equation Eq. \eqref{eq:PM} (solid line) predicts the location of DifRR (radiation \circled{1}),  the exact reason for the generation of radiation  \circled{3} is still unknown.

  \noindent \textbf{Origin of radiation \circled{3}}:	
  In our system we consider a WA with alternative gain-loss arrangement having a neutral waveguide channel  at center ($n=0$).
  	We obtain the dispersion relation for this system by analyzing the array as diatomic lattice or two-level system  \cite{xu_experimental_2016}, $\beta_{\mathcal{PT}}(\kappa) = \sqrt{4 c^2 \cos^2 \kappa - \Gamma^2}$, where the edges of the Brillouin zone is defined at $\kappa = \pm \pi/2$.
  	Adopting the PM condition similar to Eq.~\eqref{eq:PM}, we can expect a $\mathcal{PT}$-symmetry assisted radiation when, 
  	\begin{equation}
  		\beta_{\mathcal{PT}}(\kappa) - \beta_{\mathcal{PT}}(\kappa_0) - D_1 \Delta \kappa = \psi_0^2/2,
  		\label{eq:PT_PM}
  	\end{equation}
  	where $D_1 = \partial_{\kappa} \beta_{\mathcal{PT}}(\kappa)|_{\kappa = \kappa_0}$.
  The solutions to Eq.~\eqref{eq:PT_PM} are subjected to the boundary condition $\abs{\kappa} = \pi/2$ due to the resulting Brillouin zone where a shift of $\pm \pi$ in the wavenumber is required for anomalous recoiling. The proposed PM relation Eq.~\eqref{eq:PT_PM} accurately predicts the $\mathcal{PT}$ symmetry assisted resonant radiation  in $\kappa$-space as demonstrated in \figref{fig:4} (c).  We generalize the theory for a range of $k_0$ (violet solid lines) and found an excellent agreement with full numerical rasults (violet solid dots) as shown in  \figref{fig:4} (d).

\begin{figure}[h]
	\begin{center}
		\includegraphics[width=\linewidth]{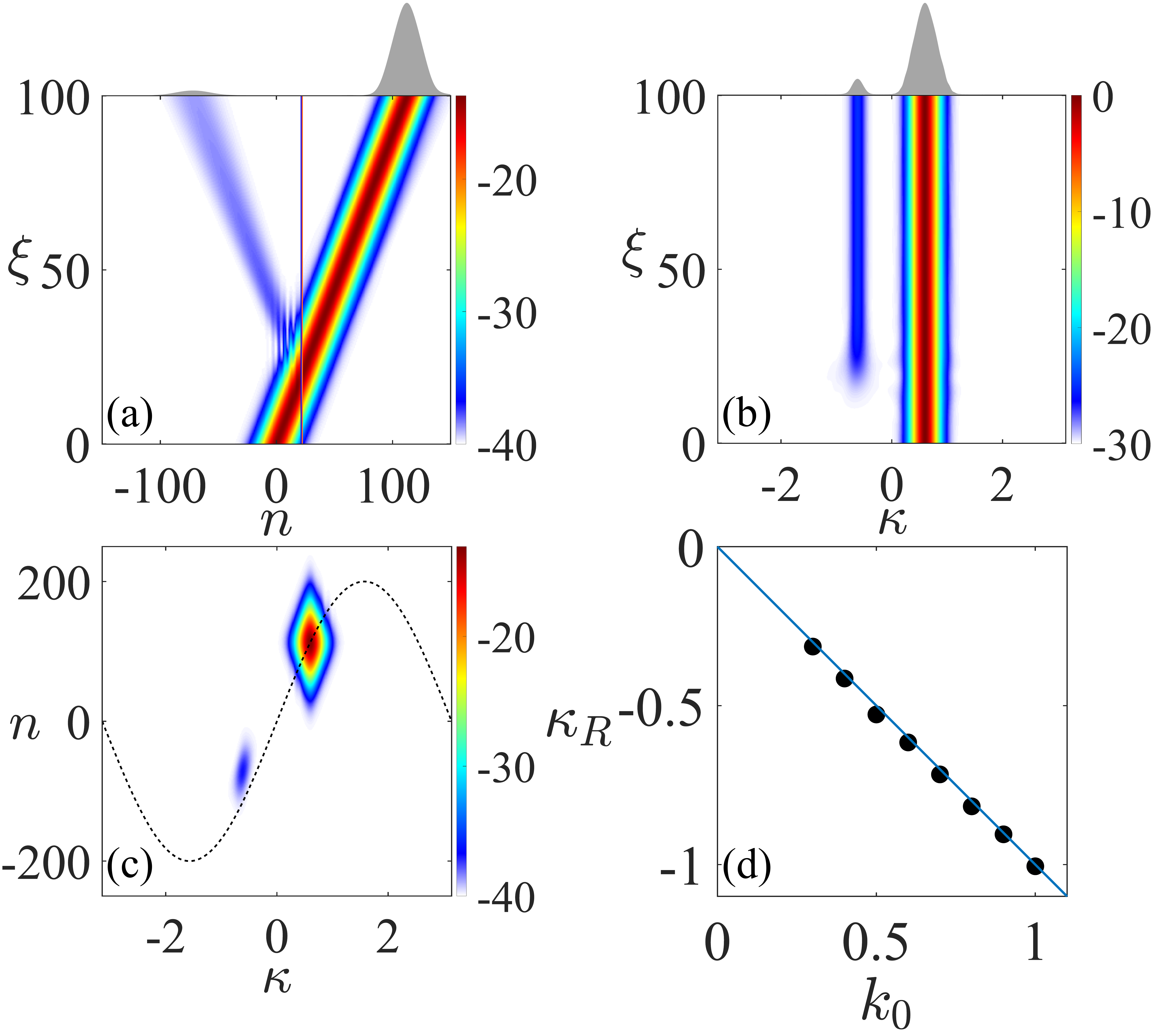}
	\end{center}
	\caption{Interaction of a low amplitude soliton of $\psi_0 = 0.2$ from a single $\mathcal{PT}$-symmetric element showing major transmission pass it, with a low amplitude reflection in the spatial domain. (b) The reflected component emerges as a radiation  at $\kappa = -k_0 = 0.6$ in the $\kappa$ domain. (c) Spectrogram at output showing both soliton and reflected component follow the delay-line. (d) Locations of the reflected component in $\kappa$-space obtained numerically which obeys the reflection condition of $\kappa = -k_0$ shown by the solid blue line.}
	\label{fig:7}
\end{figure}

\subsection{Dynamics of weak DS (for $k_0 \neq 0$) in $\mathcal{PT}$-symmetric WA}

\begin{figure}[h]
	\begin{center}
		\includegraphics[width=\linewidth]{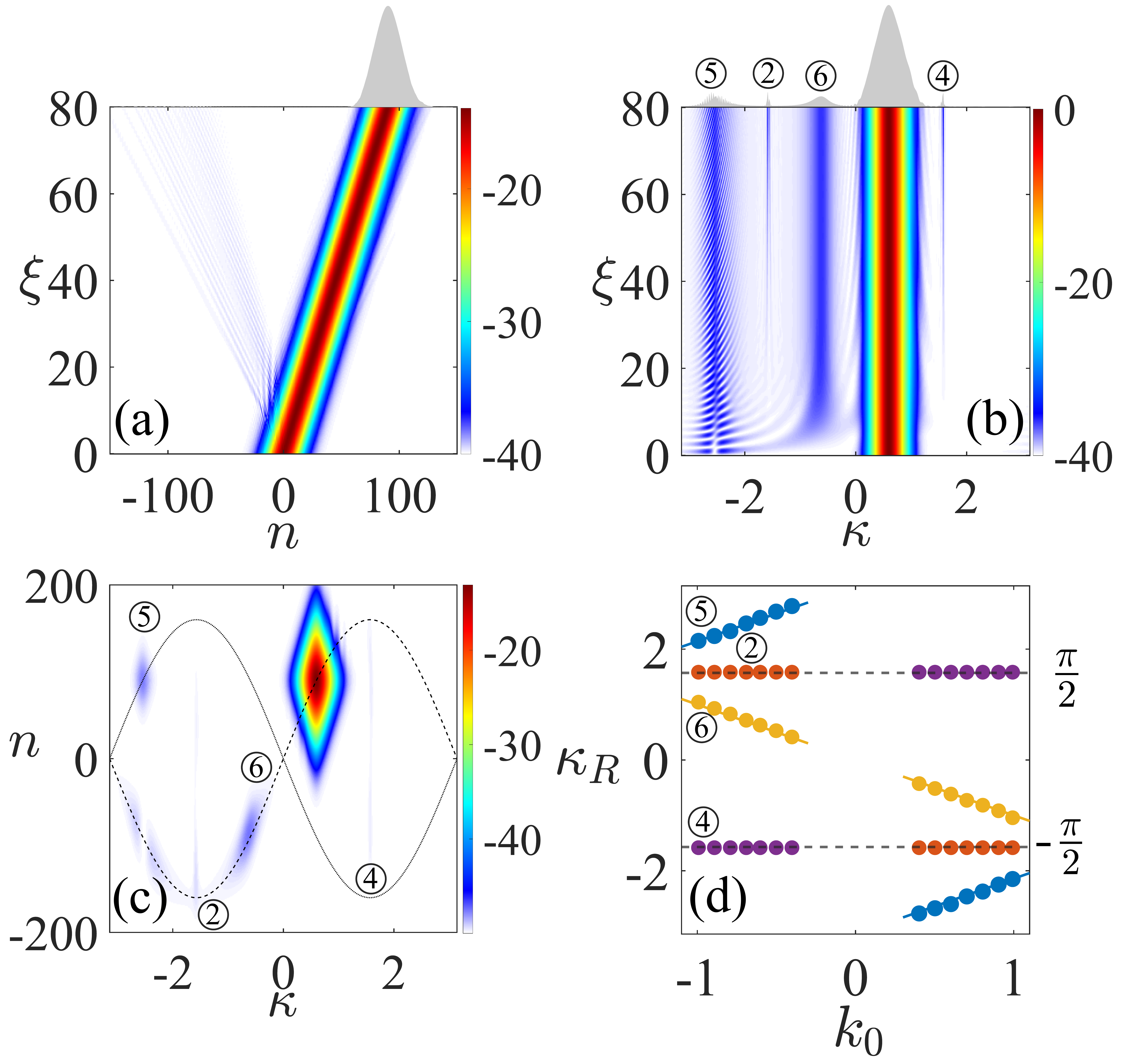}
	\end{center}
	\caption{(a) Propagation of a low amplitude ($\psi_0 = 0.2$, $k_0 = 0.6$) soliton in a $\mathcal{PT}$-symmetric WA with where DifRR is suppressed. (b) Different radiations on $\kappa$-space marked as \circled{2}, \circled{4}, \circled{5} and \circled{6}. (c) Septrogram at output with delay curves. (d) Variation of different radiations as a function of $k_0$. The solid dots are numerical results where as the lines corresponds to the theory.}
	\label{fig:8}
\end{figure}

In this subsection, we investigate the dynamics of a  weak soliton inside a $\mathcal{PT}$-symmetric WA. Here we mainly show how the radiations emitted by a weak soliton are characteristically  different from the radiation emitted by a strong DS. Based on the phase plot (see \figref{fig:2} (d) in section II B)  the soliton parameters are chosen in such a way that  both DifRR and  $\mathcal{PT}$-symmetry assisted radiation are suppressed. Note, the spatial width of the DS is inversely proportional to its amplitude and hence a weak soliton encompasses a large number of waveguide channels and also experience less amount of \textit{Peierls-Nabarro} potential \cite{Krolikowski_Soliton_1996}  during transverse propagation. This increases the interaction possibility of the soliton with periodic gain/loss channels.  Under such condition first we consider only one pair of $\mathcal{PT}$-symmetric defect channel located at $n=30,31$ that acts as a binary lattice.   In  \figref{fig:7} (a) we demonstrate the propagation of the weak DS which experiences a defect created by the pair of $\mathcal{PT}$-symmetric binary lattice. Note that due to the low amplitude and initial momentum, DS doesn't produce the regular DifRR in $\kappa$-space, the only radiation it produces is due to the reflection from defect boundary which is evident in \figref{fig:7} (b). The spectrogram plot in \figref{fig:7} (c) showcases the formation of radiation \circled{6} due to the scattering from defect boundary. A simple momentum conservation predicts the location of this radiation in $\kappa$-space as $\kappa_{R}=-k_0$ and demonstrated in  \figref{fig:7} (d) where solid dots represent the $\kappa_R$ value obtained numerically. 
Note in this system we consider a single defect crated by a balanced gain-loss channel and hence the characteristic radiations at $\pm \pi/2$ in $\kappa$-space is absent.

   In our original set-up the $\mathcal{PT}$-symmetric defects are distributed periodically throughout the waveguide array.
   When we extend our investigation in such system for weak DS, few more interesting aspects emerge which were obscured in the dynamics of a strong soliton. Here the weak soliton ($\psi_0=0.2$) with initial phase gradient radiates four distinct radiations [see \figref{fig:8} (a), (b)]. Radiation \circled{6} can be identified as refection which arises due to the successive scattering from defect boundary.  The localization of the fields \circled{2} and \circled{4} at $\pm \pi/2$ in $\kappa$-space is also explained in the earlier sections. However formation of the radiation \circled{5} is surprising and one can confuse it with the DifRR. Radiation \circled{5} must be different from DifRR since formation of DifRR is suppressed for the given parameters of the soliton.  We observe  that, a forward propagating weak soliton (which covers more waveguide channels) experiences a significant back-scattering due to the periodic $\mathcal{PT}$-symmetric defect and produces an additional radiation (in $\kappa$-space). Notably the radiation \circled{5} is originated almost from $\xi=0$ point which supports the back scattering theory. In the spectrogram plot [\figref{fig:8} (c)] also the signature of back-scattering \circled{5} is evident as  it is located at the same spatial location ($n$ coordinate)  that of the soliton but on the reverse delay curve. The phase of the back-scattered wave is detuned by $\pi$ and under folded  Brillouin zone scheme one can calculate it's momentum as $\kappa_{bs}=(k_0-\pi)$. In  \figref{fig:8} (d) we plot the momentum of all four radiations (including \circled{5}) as a function of input soliton wavenumber $k_0$ and found a satisfactory matching of simulation data with the theory that we proposed. 
    
\section{Conclusions}

In this article we explain the origin of few unique radiations that take place when a discrete soliton is excited inside a $\mathcal{PT}$-symmetric waveguide array. The concept of $\mathcal{PT}$-symmetry is optically realized by constructing a waveguide array with alternative gain-loss channels. 
We perform a linear Hamiltonian analysis to obtain the operational regime in a  $\mathcal{PT}$-symmetric system. The dynamics of a spatial soliton is investigated under such system where several distinctive radiations are emerged in the momentum space due to the $\mathcal{PT}$ symmetry. 
A strong discrete soliton when launched with an inclination  emits diffractive resonance radiation.  The $\mathcal{PT}$-symmetric nature of the waveguide array, however, excites several other radiations which were never explored earlier. We make an attempt to investigate  the physical origin of all these radiations and put forward a background theory. It is found that, under the limit of broken $\PT$ symmetry when the optical field launched normally to the central neutral channel it squeezes to the adjacent gain waveguide with pedestal extending over the $n$-space. This leads to the population of field at $\pm \pi/2$ in momentum space ($\kappa$-space) and identified as a characteristic radiation of a  $\mathcal{PT}$-symmetric waveguide array. Distinctive $\mathcal{PT}$ symmetric assisted radiation also emerges for a strong soliton due to diatomic arrangement of lattices. We establish a modified phase-matching relation that predicts the location of this radiation in momentum space.
A weak soliton also emits a few distinct radiations which are characteristically different.
For example, the alternating gain-loss channels act like a defect and the interaction between a weak soliton with these defects results radiation in the form of scattering.
In such systems, a back scattering is also observed which is prominent for a weak soliton propagating through a full $\mathcal{PT}$-symmetric system and leads to a different kind of radiation.
We characterized all these radiations and establish a background theory which is found to be  in well agreement with  full numerical simulation. Our results might be useful for next generation light shaping and optical switching applications.

\section*{ACKNOWLEDGMENT}
	A.P.L. acknowledges University Grants Commission, India for support through Junior Research Fellowship in Sciences,	Humanities and Social Sciences (ID 515364)
	
		
	\bibliography{refs.bib}
	\bibliographystyle{apsrev4-2}
\end{document}